\documentclass[twocolumn,showpacs,superscriptaddress,preprintnumbers,amsmath,amssymb]{revtex4}

\usepackage{graphicx}	
\usepackage{dcolumn}	
\usepackage{bm}		
\usepackage{color}	
\usepackage{multirow}	
\usepackage{longtable,epsfig}
\usepackage{sidecap}

\begin{document}

\title{Multiple chiral doublet bands of identical configuration in $^{103}$Rh}

\author{I.~Kuti}
\affiliation{Institute for Nuclear Research, Hungarian Academy of Sciences, Pf. 51, 4001 Debrecen, Hungary}

\author{Q.~B.~Chen}
\affiliation{State Key Laboratory of Physics and Technology, School of Physics, Peking University, Beijing 100871, China}

\author{J.~Tim\'ar}
\affiliation{Institute for Nuclear Research, Hungarian Academy of Sciences, Pf. 51, 4001 Debrecen, Hungary}

\author{D.~Sohler}
\affiliation{Institute for Nuclear Research, Hungarian Academy of Sciences, Pf. 51, 4001 Debrecen, Hungary}

\author{S.~Q.~Zhang}
\affiliation{State Key Laboratory of Physics and Technology, School of Physics, Peking University, Beijing 100871, China}

\author{Z.~H.~Zhang}
\affiliation{State Key Laboratory of Physics and Technology, School of Physics, Peking University, Beijing 100871, China}

\author{P.~W.~Zhao}
\affiliation{State Key Laboratory of Physics and Technology, School of Physics, Peking University, Beijing 100871, China}

\author{J.~Meng}
\affiliation{State Key Laboratory of Physics and Technology, School of Physics, Peking University, Beijing 100871, China}

\author{K.~Starosta}
\affiliation{Department of Chemistry, Simon Fraser University, Burnaby, British Columbia V5A 1S6, Canada}

\author{T.~Koike}
\affiliation{Graduate School of Science, Tohoku University, Sendai, 980-8578, Japan}

\author{E.~S.~Paul}
\affiliation{Oliver Lodge Laboratory, University of Liverpool, Liverpool L69 7ZE, United Kingdom}

\author{D.~B.~Fossan}
\affiliation{Department of Physics and Astronomy, State University of New York, Stony Brook,~New York,~11794-3800, USA}

\author{C.~Vaman}
\affiliation{Department of Physics and Astronomy, State University of New York, Stony Brook,~New York,~11794-3800, USA}

\date{\today}

\begin{abstract}
Three sets of chiral doublet band structures have been identified in the $^{103}$Rh nucleus.
The properties of the observed chiral doublet bands are in good agreement with theoretical 
results obtained using constrained covariant density functional theory and particle rotor model calculations.
Two of them belong to an identical configuration, and provide the first experimental evidence for
a novel type of multiple chiral doublets, where an ``excited'' chiral doublet of a configuration
is seen together with the ``yrast'' one.  This observation shows that the chiral geometry in nuclei
can be robust against the increase of the intrinsic excitation energy.
\end{abstract}

\pacs{21.10.Hw,21.10.Re,21.60.-n,23.20.Lv}

\maketitle


A novel form of spontaneous symmetry breaking, the chiral rotation of triaxial 
nuclei, was suggested in 1997 \cite{1}. It was shown that in special circumstances, 
referred to as chiral geometry, in the intrinsic frame of the rotating triaxial
nucleus the total angular momentum vector lies outside the three principal
planes. Thus, its components along the principal axes can be oriented in
left- and right-handed ways. In the laboratory frame the chiral symmetry
is restored, which manifests itself as a pair of $\Delta I=1$ nearly
degenerate bands with the same parity. Such chiral doublet bands were
first identified in four $N = 75$ isotones in 2001~\cite{2}. So far, many
chiral candidate nuclei have been reported experimentally in the $A\sim$ 80,
100, 130, and 190 mass regions~\cite{he2,ha3,ko4,bark,star2,ko6,ra7,zhu1,jain,
srebrny,gr9,vam1,pank1,pank2,alcan1,tim1,tim2,a190,A80}. 
Besides the simplest chiral configurations composed of one unpaired proton and 
neutron, composite chiral configurations, containing more than one unpaired 
protons and/or neutrons, have also been observed in the odd-mass or even-even 
neighbors of the odd-odd chiral nuclei~\cite{zhu1,tim1}. These observations show that 
chirality is not restricted to a certain configuration in a mass region, i.e. the 
chiral geometry can be robust against the change of configuration.
It was even demonstrated recently by Meng \emph{et al.}~\cite{6,7,8,9}, 
based on adiabatic and configuration-fixed constrained triaxial covariant density 
functional theory (CDFT) calculations,
that it is possible to have multiple pairs of chiral doublet bands in a single nucleus,
and the acronym M${\chi}$D was introduced for this phenomenon.
The first experimental evidence for the predicted M${\chi}$D has been reported 
in $^{133}$Ce \cite{10}, and also possibly in $^{107}$Ag~\cite{11}. 

It is also interesting to study the robustness of chiral geometry against the increase 
of the intrinsic excitation energy, i.e. if the chiral geometry is sustained in the higher-lying 
bands of a certain chiral configuration. In all the known cases the chiral doublet corresponds 
to the two lowest-lying bands of a configuration. Even for M$\chi$D in $^{133}$Ce~\cite{10} 
and $^{107}$Ag~\cite{11}, each chiral doublet structure corresponds to two lowest-lying 
bands with a distinct configuration. Therefore, study of the third and forth 
bands of the same chiral configuration is needed to answer the question of the 
investigated robustness. 
Very recent model calculations predicted multiple chiral doublet bands which belong 
to the same configuration~\cite{25,26,27}. In this Letter we report on the first 
experimental evidence for such a type of M${\chi}$D in the $^{103}$Rh nucleus.


\begin{figure*}
\includegraphics[width=120mm,angle=-90,bb=60 150 530 650]{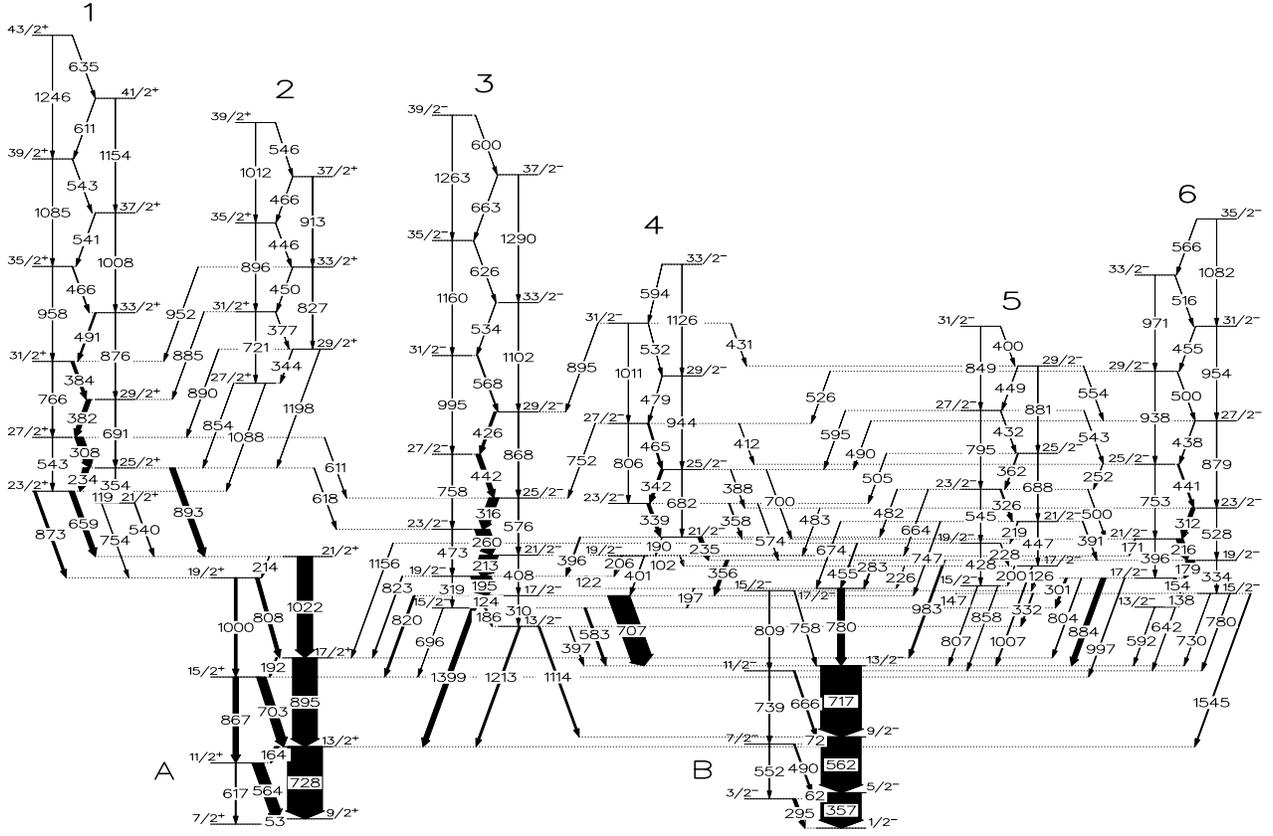}
\caption{Partial level scheme of $^{103}$Rh. The energies are given in keV, the
widths of the arrows are proportional to the relative transition intensities.}
\label{lslev}
\end{figure*}

\begin{figure}
\includegraphics[width=40mm,angle=-90,bb=0 30 410 900]{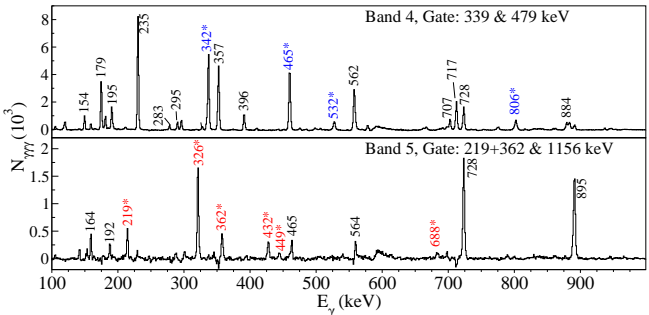}
\caption{(Color online) Typical $\gamma\gamma\gamma$-coincidence spectra obtained in the present work
showing the placement of the $\gamma$ rays in bands~4 and 5. Transitions marked with star and labeled
in blue or red are inband transitions.}
\label{spectra}
\end{figure}

Medium- and high-spin states of $^{103}$Rh were populated using the
$^{96}$Zr($^{11}$B,4n) reaction at a beam energy of 40 MeV. The beam, provided
by the 88-inch Cyclotron of the LBNL, impinged upon an enriched 500 $\mu$g/cm$^2$
thick self-supporting Zr foil. The emitted $\gamma$-rays were detected by the
Gammasphere spectrometer. Approximately $9\times 10^8$ four- and higher-fold events
were accumulated and sorted off-line into 2-d and 3-d histograms. The data analysis
was carried out using the {\sc radware} software package~\cite{RW}. A more complete
level scheme of $^{103}$Rh was constructed using the observed coincidence relations
and relative intensities of the gamma transitions and based on the formerly reported
states~\cite{tim2,21}. 
Spin assignments for the new states were deduced from the measurements of 
angular-intensity ratios, based on the method of directional correlation from oriented states 
(DCO)~\cite{DCO}. The parities were deduced using the additional assumption that if a 
level decays to a band by both quadrupole and dipole transitions with comparable intensities, then 
the quadrupole transition is E2 and the dipole is M1.
Several new rotational bands have been found and the previously reported ones were 
extended to higher spins. A partial level scheme showing the bands relevant to the focus of 
this Letter is plotted in Fig.~\ref{lslev}. Bands A and B were formerly reported in Ref.~\cite{21}, 
while bands 1 and 2 were identified earlier and assigned as a chiral doublet with
configuration $\pi(g_{9/2})\otimes \nu(h_{11/2})^2$ in Ref.~\cite{tim2}. Bands 3 and 6
were first reported in Ref.~\cite{21} up to spins 29/2 and 25/2, respectively,
with a tentative configuration assignment of ${\pi}(g_{9/2})^{2}(p_{1/2})$. In the
present work we extended these bands to spins 39/2 and 35/2, respectively. 
Two of the newly observed bands, labeled as band 4 and 5, are presented in Fig.~\ref{lslev}.
Fig.~\ref{spectra} shows triple-coincidence $\gamma$-ray spectra proving the placements
of the levels in bands 4 and 5, with the in-band transitions being highlighted. For these
bands the spin-parities were deduced from DCO measurements. In the present geometry,
setting the gate on a stretched quadrupole transition, {\sl R}$_{\rm DCO}$ values of
$\sim$1.0 and $\sim$0.5 were expected for stretched quadrupole and stretched dipole
transitions, respectively. An $E2$ electromagnetic character to the 674- and 1007~keV
transitions linking band 5 to band B was assigned based on the DCO values of 0.93(11) and
1.07(4), respectively, setting the gate on the 717~keV $E2$ transition of band B.
Similarly the 574- and 700~keV transitions linking band 4 to band 6 are considered to have
$E2$ character, based on the DCO values of 0.89(18) and 0.88(19) obtained by gating on
the 884~keV $E2$
transition linking band 6 to band B. These quadrupole transitions fix the negative parity and the
spins for bands 4 and 5. This assignment is verified by the 0.55(2) and 0.54(13) DCO values
of the 396- and 391~keV stretched $M1$ transitions linking band 4 to band 3 and band 5 to 
band 6, respectively.

\begin{figure}
\includegraphics[width=85mm,bb=20 30 775 362]{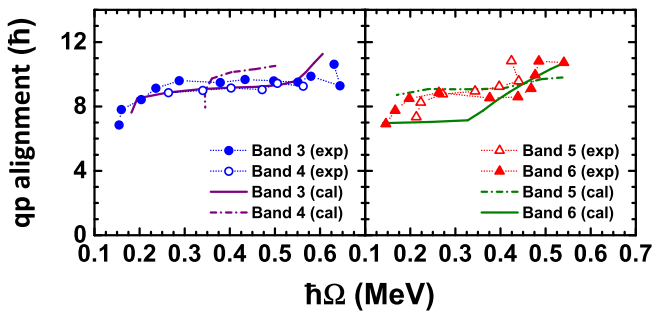}
\caption{(Color online) Quasiparticle alignments of bands~3,4,5,6. For the details, see text.}
\label{alig}
\end{figure}

The $\Delta I=1$ bands 3-6 all have negative parity and are linked to each other
by many transitions. Furthermore, the energy differences between the same-spin
levels are rather small, for bands 3 and 4 these values are about 300 keV and for 
bands 6 and 5 about 100 keV, and their {\sl B(M1)/B(E2)} values are very similar. 
These properties may indicate chirality. Indeed, in neighboring $^{105}$Rh
similar structure of three $\Delta I=1$ bands have been observed and identified
tentatively as the three lowest-energy bands of the ${\pi}g_{9/2}{\otimes}{\nu}h_{11/2}(g_{7/2},d_{5/2})$
configuration \cite{alcan1}. In $^{105}$Rh the lowest-energy one of the three bands
is thought to have planar geometry, while the two higher-energy bands are considered
as a chiral doublet. In $^{103}$Rh, we see four bands which could correspond to
two chiral doublets. To assist in the comparison of the properties of bands 3-6,
we have derived their quasiparticle alignments as defined in Ref.~\cite{bengtsson}.
$K=1/2$ and the $\mathcal{J}_0=7~{\hbar}^2/$MeV and $\mathcal{J}_1=15.7~{\hbar}^4/$MeV$^{3}$ 
parameters of the Harris formula $\mathcal{J}=\mathcal{J}_0+\mathcal{J}_1\omega^2$, describing the 
dependence of moments of inertia on the rotational frequency, have been adopted in the derivation. The
obtained alignments are shown in Fig.~\ref{alig}. Although the alignment value is
${\sim}9{\hbar}$ for all four bands in a wide interval of rotational frequency,
there is a pronounced similarity between bands 3 and 4, and also between bands 5 and 6.
These similarities enable us to group the four bands into two possible chiral doublets.
According to this grouping the lowest-energy and the second lowest-energy bands
(bands 3 and 4, respectively) could form the ``yrast'' chiral doublet, while the next
two bands in energy (bands 6 and 5) could form the ``excited'' chiral doublet of the
probable ${\pi}g_{9/2}{\otimes}{\nu}h_{11/2}(g_{7/2},d_{5/2})$ configuration. We need
to mention, however, that bands 4 and 6 are so close to each other in energy that their
energy ordering varies with spin.


In order to understand the nature of the observed band structure in $^{103}$Rh,
first adiabatic and configuration-fixed constrained CDFT calculations~\cite{6} were performed
to search for the possible configurations and deformations. Subsequently, the configurations and
deformations were further confirmed and reexamined by tilted axis cranking 
CDFT (TAC-CDFT) calculations \cite{13,14,15,16} determining the energy spectra, Routhians,
spin-frequency relations, deformations, and alignments. Finally, with the obtained configurations
and deformations, quantum particle rotor model \cite{10,17,18} calculations were performed
to study the energy spectra and {\sl B(M1)/B(E2)} ratios for both the positive- and the
negative-parity bands.

The potential-energy surface in the $\beta$-$\gamma$ plane obtained from the CDFT calculations
with effective interaction PC-PK1~\cite{19} shows that the ground state of $^{103}$Rh has a triaxial 
deformation with a quadrupole deformation of $\beta = 0.25$ and a triaxiality parameter of 
$\gamma =20^\circ$, and is soft with respect to the $\gamma$ degree of freedom.
The configuration-fixed calculations provided the ${\pi}(1g_{9/2})^{-1}$ 
(more precisely, seven protons in $g_{9/2}$ shell) and the ${\pi}(2p_{1/2})^{1}$
unpaired-nucleon configurations for bands A and B, in a good agreement with previous configuration
assignments~\cite{21}. Among the five lowest-lying configurations with three unpaired nucleons,
one positive-parity and four negative-parity configurations have been found. The band-head of
the positive-parity ${\pi}(1g_{9/2})^{-1}{\otimes}{\nu}(1h_{11/2})^{2}$ configuration is predicted
to have a triaxial shape of $\beta=0.29$, $\gamma=11.0^{\circ}$. Two of the four negative-parity
bands are predicted to have triaxial shape, namely configuration $\pi(1g_{9/2})^{-1}\otimes
\nu(1h_{11/2})^1(2d_{5/2})^1$ with $\beta=0.27$, $\gamma=18.7^\circ$ and configuration  
$\pi(1g_{9/2})^{-1}\otimes\nu(1h_{11/2})^1(1g_{7/2})^{-1}$ with $\beta=0.26$, $\gamma=14.5^\circ$.
Here $(g_{7/2})^{-1}$ denotes the occupation of five neutrons in the $g_{7/2}$ shell.

In order to study the rotational behavior of the predicted configurations, and to
reexamine their configurations and deformations with rotation, TAC-CDFT~\cite{13,14,15,16}
with the effective interaction PC-PK1~\cite{19} was adopted. Due to the strong mixing between low-$j$ orbits,
we kept the configurations of the high-$j$ valence nucleons fixed and left the other nucleons automatically
occupying the lowest levels. For positive-parity bands, we fixed the high-$j$ orbitals
${\pi}(1g_{9/2})^{-1}{\otimes}{\nu}(1h_{11/2})^{2}$, while for negative-parity bands
we fixed the ${\pi}(1g_{9/2})^{-1}{\otimes}{\nu}(1h_{11/2})^{1}$. The calculated energies as a
function of spin, as well as the Routhians and spins as a function of the rotational
frequency for the above two configurations are in a good agreement with the experimental
values for bands 1,2 and 3-6, respectively.  With increasing rotational frequency, the
$\beta$ shape parameter was found to decrease somewhat, while the $\gamma$ parameter
increased to around 30$^{\circ}$ with an average values of 20$^{\circ}$ in the observed
frequency range. For the low-$j$ components of the negative-parity configuration, a $g_{7/2}$ 
neutron is found to contribute a large alignment along short axis (about $3\hbar$) in the TAC-CDFT 
calculations and the contribution from other $g_{7/2}$ neutrons can be regarded as a part of core. 
Hence the valence $g_{7/2}$ neutron contributions can be approximated by an effective $(1g_{7/2})^1$ 
configuration. Therefore in the following discussions, the configuration for the negative parity bands 
is written as $\pi g_{9/2}^{-1} \otimes \nu h_{11/2} g_{7/2}^{1}$.
It has also been revealed by the calculations that the angular momenta of the
$(1h_{11/2})^{1}$ and $(1g_{7/2})^{1}$ neutrons are aligned along the short axis,
while the angular momentum of the $(1g_{9/2})^{-1}$ proton is mainly aligned along
the long axis, as it is expected in case of chirality. The details will be presented
in a forthcoming publication.

\begin{SCfigure*}
\caption{(Color online) Experimental excitation energies and {\sl B(M1)/B(E2)} ratios
for the positive-parity chiral bands 1-2 (left panels) and negative-parity multiple chiral
bands 3-6 (middle and right panels) in $^{103}$Rh together with the results of triaxial particle rotor
model. The number following the configuration label of the theoretical curve corresponds to the
energy ordering of the calculated band with the given configuration.}
\includegraphics[width=0.7 \textwidth]{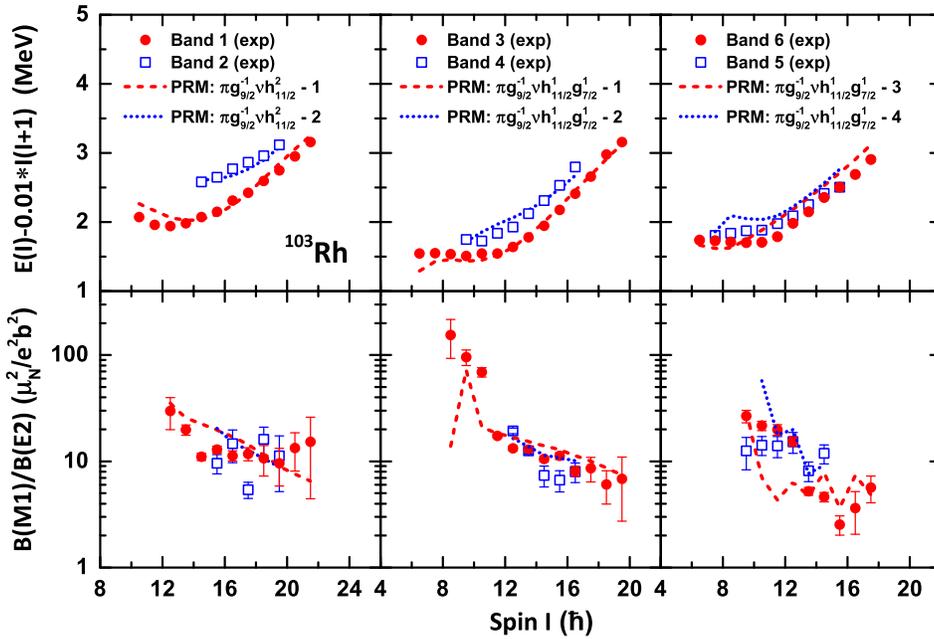}
\label{spec}
\end{SCfigure*}

In order to study the energy spectra and the {\sl B(M1)/B(E2)} ratios of the positive-
and negative-parity bands, the quantum particle rotor model (PRM)~\cite{10,17,18} has
been applied for both the ${\pi}(1g_{9/2})^{-1}{\otimes}{\nu}(1h_{11/2})^{2}$ and
${\pi}(1g_{9/2})^{-1}{\otimes}{\nu}(1h_{11/2})^{1}(1g_{7/2})^{1}$ configurations.
In the PRM calculations, the input $\beta$ deformation parameter at the bandhead was
$\beta=0.29$ for the positive parity bands and $\beta=0.26$ for the negative parity bands. 
The $\gamma$ parameter was adopted as 20$^{\circ}$, which was obtained from the 
TAC-CDFT calculations. The single-$j$ shell Hamiltonian parameter was taken as 
$C=\left(\frac{123}{8}\sqrt\frac{5}{\pi}\right)\frac{2N+3}{j(j+1)}A^{-1/3}\beta$~\cite{23}.
For the electromagnetic transitions, the empirical intrinsic quadrupole moment of
$Q_0 = (3/\sqrt{5\pi})R_0^2 Z\beta$ with $R_0=1.2A^{1/3}$~fm and the gyromagnetic ratios of
$g_R= Z/A$, $g_\pi(g_{9/2}) = 1.26$, $g_\nu(h_{11/2}) =-0.21$, $g_\nu(g_{7/2})=0.70$
were adopted.

For the two positive-parity bands with the configuration ${\pi}(1g_{9/2})^{-1}
{\otimes}{\nu}(1h_{11/2})^{2}$, a moment of inertia $\mathcal{J}_0 = 23~\hbar^2$/MeV
was used. This was adjusted to reproduce the trend of the energy spectra of bands 1 and 2.
The obtained energy spectra are shown in Fig.~\ref{spec}. The PRM results excellently
agree with the data. These two bands are separated by $\sim$500 keV at $I =29/2~\hbar$.
They approach each other with increasing spin and the separation finally goes to $\sim 360$ keV at
$I=39/2$. The $B(M 1)/B(E2)$ values of bands 1 and 2 are similar. 
The observation that the experimental $B(M1)/B(E2)$ values for bands 1 and 2 do not fall off as 
quickly with spin as the theoretical values comes from the frozen rotor assumption adopted in PRM.
In Ref.~\cite{18}, a detailed analysis shows that in both $^{103}$Rh and
$^{105}$Rh, the chiral bands with positive parity change from chiral vibration to nearly static 
chirality at spin $I = 37/2$ and back to another type of chiral vibration at higher spins. 
Such a conclusion is still held here for the positive-parity doublet.

For the four negative-parity bands of configuration ${\pi}(1g_{9/2})^{-1}{\otimes}
{\nu}(1h_{11/2})^{1}(1g_{7/2})^{1}$, a moment of inertia of $\mathcal{J}_0 = 25~\hbar^2$/MeV
was adopted. A Coriolis attenuation factor of $\xi$ = 0.85 has been employed to take into
account the effect of the strong mixing between low-$j$ neutrons. In Fig.~\ref{spec}, the
four lowest-energy calculated bands of the above configuration are compared with the
experimental bands 3-6. The four calculated bands form two chiral doublets, of which
the first one fits the experimental band-pair 3 and 4, while the second doublet can also 
reasonably reproduce the trend of bands 6 and 5. The calculated energies for bands 5 
and 6 are higher than the experimental values about 200 keV, which might be ascribed to 
that the complex correlations are not fully taken into account in the PRM calculations with 
single-$j$ shell Hamiltonian.
The corresponding calculated electromagnetic transition probabilities, shown in Fig.~\ref{spec}, are 
also able to reproduce the data reasonably. The weak odd-even $B(M1)/B(E2)$ staggering 
for bands 3 and 4 is consistent with the case of chiral vibration as discussed in Ref.~\cite{24}. 
For bands 5 and 6, the $B(M1)/B(E2)$ values show a staggering at $I=15.5~\hbar$, which is 
also reproduced by the PRM.

In contrast with the multiple chiral doublets predicted in Ref.~\cite{6} and experimentally
reported in $^{133}$Ce~\cite{10}, the observed M${\chi}$D in the negative-parity bands
of $^{103}$Rh is built from the first and second doublets of the same configuration.
Although the two doublets belong to the same configuration, the angular momenta couplings 
for the two pairs of chiral partners are different. This fact is reflected by the different alignment
properties of the two doublets.
The quasiparticle alignments are extracted from the calculated energy spectra by using the same 
experimental Harris parameters for bands 3-6. It can be seen that the calculated alignments to 
some extent are in agreement with the experimental values. The sharp increase of alignments around 
$\hbar\omega= 0.45$ MeV for bands 5, 6 are not reproduced, which might be attributed to a 
frozen core used in the framework of particle rotor model. The theoretical alignment shows a 
smooth increasing rather than a sharp increasing.
Observation of M${\chi}$D with the same configuration shows that the chiral geometry in nuclei can be
robust against the increase of the intrinsic excitation energy.


In summary, one positive-parity and two negative-parity chiral doublet band structures
have been identified in $^{103}$Rh. The observed doublet bands have been compared with
results of calculations involving adiabatic and configuration-fixed constrained CDFT,
TAC-CDFT, and quantum particle rotor model.
The theoretical results reproduce the data rather well. According to these results,
the positive-parity doublet has a chiral vibrational structure based on the ${\pi}(1g_{9/2})^{-1}
{\otimes}{\nu}(1h_{11/2})^{2}$ configuration, while the two negative-parity doublets
are chiral doublets with the same ${\pi}(1g_{9/2})^{-1}{\otimes}{\nu}(1h_{11/2})^{1}(1g_{7/2})^{1}$
configuration. It provides the first experimental evidence for the M${\chi}$D with the same configuration, and 
shows that chiral geometry can be robust against the increase of the intrinsic excitation energy.

The crew and staff of the 88-Inch Cyclotron are thanked. Special thanks to A.~O.~Macchiavelli and
I.~Y.~Lee for their help in the experiment.
This work was supported in part by the Hungarian Scientific Research Fund, OTKA (Contract No. K100835),
the Major State 973 Program of China (Grant No. 2013CB834400), the  National Natural Science Foundation
of China (Grants No. 11175002, No. 11335002, No. 11375015), Research Fund for the Doctoral Program of
Higer Eductation (Grant No. 20110001110087), the China Postdoctoral Science Foundation (Grant No. 2012M520101
and No. 2013M540011), the Natural Sciences and Engineering Research Council of Canada under Contract
No. SAPIN/371656-2010 and the UK Engineering and Physical Sciences Research Council.


\end{document}